\documentclass[aip, jcp, notitlepage, floatfix, reprint,
  citeautoscript, twocolumn]{revtex4-1}

\usepackage{amsmath}
\usepackage{amsfonts}
\usepackage{graphicx}
\usepackage{dcolumn}
\usepackage{miller}
\usepackage[table]{xcolor}
\usepackage[version=3]{mhchem}
\usepackage{transparent}
\usepackage{gensymb}
\usepackage{tabularx}
\usepackage{etoolbox}
\usepackage[hidelinks]{hyperref}
\usepackage{bm}
\usepackage{relsize}
\usepackage[normalem]{ulem}

\def\maxfloatwidth{%
  \ifdim\columnwidth>246.0pt
  300.0pt  \else
  \columnwidth
  \fi
}

\newcommand{\tbf}[1]{\textbf{#1}}

\newcommand{\tcr}[1]{\textcolor{black}{#1}}

\newcommand{\etal}{\emph{et al.}}

\begin{document}

\title{Macroscopic surface charges from microscopic simulations}

\author{Thomas Sayer}

\author{Stephen J. Cox} 
\affiliation{Department of Chemistry, University of Cambridge,
  Lensfield Road, Cambridge CB2 1EW, United Kingdom}
\email{sjc236@cam.ac.uk}

\date{\today}

\begin{abstract}
  Attaining accurate average structural properties in a molecular
  simulation should be considered a prerequisite if one aims to elicit
  meaningful insights into a system's behavior. For charged surfaces
  in contact with an electrolyte solution, an obvious example is the
  density profile of ions along the direction normal to the
  surface. Here we demonstrate that, in the slab geometry typically
  used in simulations, imposing an electric displacement field $D$
  determines the integrated surface charge density of adsorbed ions at
  charged interfaces. This allows us to obtain macroscopic surface
  charge densities irrespective of the slab thickness used in our
  simulations. We also show that the commonly used Yeh-Berkowitz
  method and the `mirrored slab' geometry both impose vanishing
  integrated surface charge density. We present results both for
  relatively simple rocksalt \hkl(111) interfaces, and the more
  complex case of kaolinite's basal faces in contact with aqueous
  electrolyte solution.
\end{abstract}

\maketitle

\section{Introduction}
\label{sec:intro}

Charged surfaces in contact with solution are commonplace in fields as
diverse as colloid science, geology and energy materials
\cite{swartzen1974surface,sposito1999surface,xu2004nonaqueous,xu2014electrolytes,salanne2016efficient,mora2017polar,hartkamp2018measuring}. As
such, there is great interest in using molecular simulations to probe
the details of these systems at the microscopic scale. However, the
long-ranged nature of electrostatic interactions, and the relatively
small system sizes typically afforded by molecular simulations can
have severe consequences for simulated observables
\cite{HummerGarcia1996sjc,YehBerkowitz1999sjc,HunenbergerMcCammon1999sjc,cox2018interfacial,zhang2016finite}. The
purpose of this article is to demonstrate how commonly used simulation
approaches lead to \emph{qualitatively} incorrect descriptions of ion
adsorption at charged interfaces. We will also extend the ideas of
previous works
\cite{zhang2016finite,sayer2017charge,sayer2019finite,sayer2019stabilization}
to not only correct for small system sizes, but to understand why
other methods fail in a dramatic fashion. In fact, this simply amounts
to setting the electrostatic boundary conditions appropriately; as
this is relatively straightforward to do in existing simulation
packages,\footnote{Source code that implements the finite field
  approach in LAMMPS is freely available at
  \url{https://github.com/uccasco/FiniteFields}.} it is hoped that the
results presented here---along with those in
Refs.~\onlinecite{zhang2016finite,sayer2017charge,sayer2019finite,sayer2019stabilization}---will
be useful to the simulation community in modeling charged solid/liquid
interfaces.

\begin{figure*}[tb]
  \centering
  \includegraphics[width=17cm]{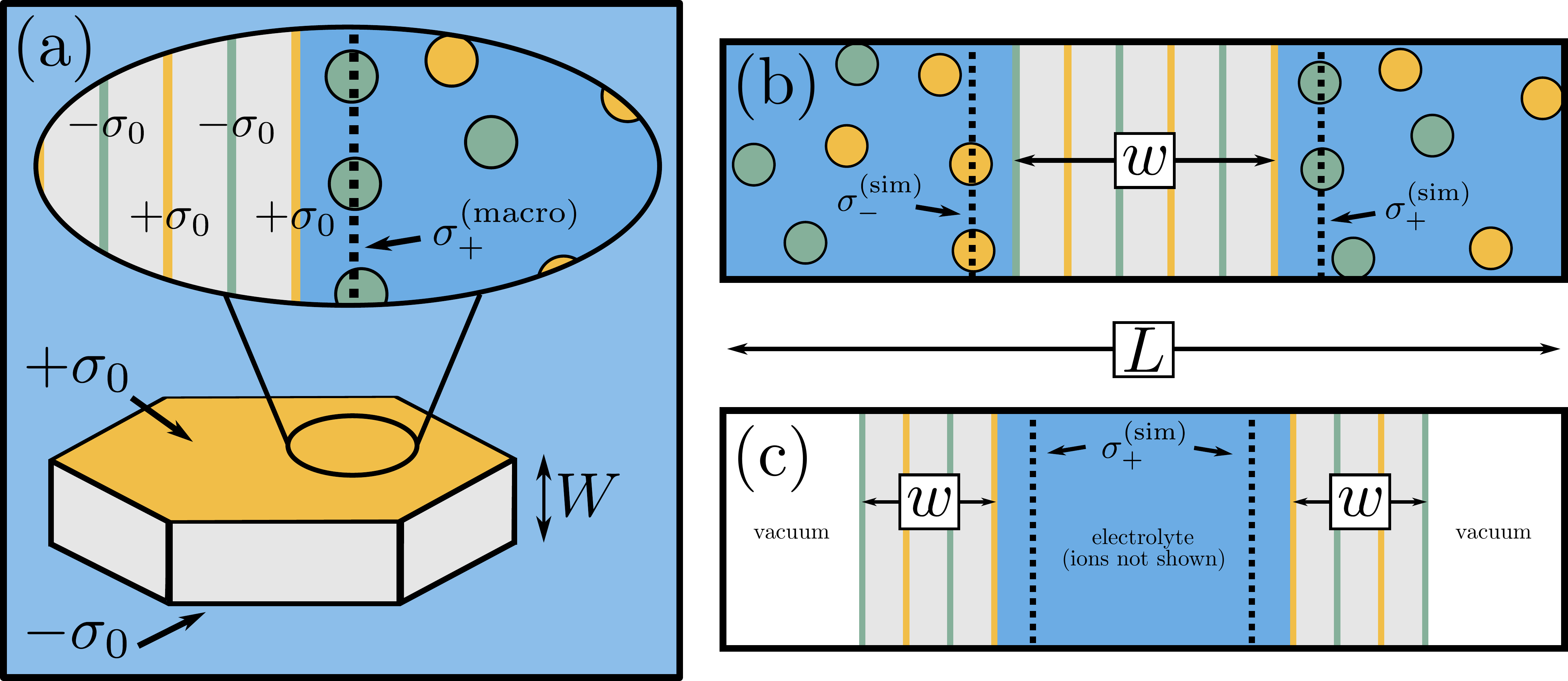}
  \caption{Schematic of polarity compensation from the solution
    environment. (a) The system of interest is a single crystal that
    predominantly exposes polar crystal facets with surface charge
    density $\pm\sigma_0$. The thickness of the crystal $W$ is
    macroscopic in extent. In the absence of other mechanisms,
    counterions will adsorb from solution to stabilize the crystal
    such that the integrated surface charge density is
    $\sigma_{\pm}^{\rm (macro)}$ (see Eq.~\ref{eqn:sigma-macro}), as
    represented by the dotted line. (b) Simulation cell employing the
    \emph{slab geometry} often used in simulations to study
    interfacial systems. The simulation cell is periodically
    replicated in all three dimensions. The thickness of the slab $w$
    is much smaller than $W$. The integrated surface charge density
    $\sigma_\pm^{\rm (sim)}$ will depend upon the electrostatic
    boundary conditions employed. (c) Simulation cell for the
    \emph{mirrored slab geometry}. Here two slabs with dipole moments
    pointing in opposite directions confine an electrolyte
    solution. Periodic images are separated by vacuum. The simulation
    cell is periodically replicated in all three dimensions. In both
    (b) and (c), $L$ denotes the length of the simulation cell in the
    direction perpendicular to the surface plane, which is taken to be
    $z$.}
  \label{fig:schematic}
\end{figure*}

To illustrate one of the main challenges faced when simulating charged
interfacial systems, it is perhaps useful to first discuss what the
physical scenario is that we aim to describe. For simplicity, we only
explicitly consider surface charge originating from polar
crystallographic axes, although it is important to note that other
mechanisms are possible e.g. protonation/deprotonation of functional
groups; our results are directly relevant to such cases too. To this
end, consider the situation depicted in
Fig.~\ref{fig:schematic}\,(a). Here, a macroscopic single crystal
exposing polar facets---resulting from the termination of the crystal
along a crystallographic direction comprising alternating planes of
opposite charge---is immersed in an electrolyte solution. It is well
established that such polar crystal terminations are inherently
unstable, and require a polarity compensation mechanism
\cite{tasker1979stability,nosker1970polar,goniakowski2008polarity,noguera2000polar,sayer2019stabilization}. In
this article, we will focus on the case where adsorption of charge
from the external environment stabilizes the crystal. Specifically, we
expect counterions from solution to adsorb to the crystal's surfaces
such that the integrated charge density over an interfacial region,
e.g.,
\begin{equation}
  \label{eqn:sigma-macro}
  \sigma_+^{\rm (macro)} = \int_{{\rm int}+}\!{\rm d}z\,n(z),
\end{equation}
provides the appropriate polarity compensation. In
Eq.~\ref{eqn:sigma-macro}, $n$ is the charge density profile
perpendicular to the interface, which we take to define the $z$
direction, and the integration is understood to be taken over the
interfacial region corresponding to the positively charged crystal
surface. A similar definition holds for $\sigma_-^{\rm (macro)}$. The
exact value of $\sigma_\pm^{\rm (macro)}$ depends upon the in-plane
charge density $\sigma_0$ and the crystal structure (see
Fig.~\ref{fig:schematic}). For example, in the case of rocksalt
\hkl(111), $\sigma_\pm^{\rm (macro)} \approx \mp\sigma_0/2$, while for
the \hkl(0001) surfaces of wurtzite isomorphs, $\sigma_\pm^{\rm
  (macro)} \approx \mp\sigma_0/4$. A reasonable goal for a molecular
simulation is to obtain an integrated interfacial charge density
$\sigma_\pm^{\rm (sim)} \approx \sigma_\pm^{\rm (macro)}$ that well
approximates the macroscopic sample of interest.

A typical classical molecular dynamics simulation comprises $\sim
10^2$--$10^5$ molecules. This is obviously far lower than what is
found in the macroscopic sample sizes that experiments can probe. To
avoid artificially large surface-to-volume ratios or degrees of
interfacial curvature, \emph{periodic boundary conditions} (PBC) are
typically applied, in which the system is periodically replicated in
all three dimensions \cite{FrenkelSmit2002sjc,allen1987computer}. Two
typical geometries of a simulation cell used to study interfacial
systems under PBC are shown in Fig.~\ref{fig:schematic}. The first of
these, shown in Fig~\ref{fig:schematic}\,(b), and simply referred to
as the \emph{slab geometry}, consists of a single slab of solid
material with thickness $w$ centered at $z=0$ and surrounded on either
side by electrolyte solution. The slab itself comprises alternating
planes of opposite charge, which at present are simply taken to have
an equidistant spacing; a simple generalization to more complex
scenarios will be presented in Sec.~\ref{sec:kao}. The total extent of
the simulation cell along $z$ is $L$, with the boundaries of the cell
at $z=\pm L/2$. We stress that we have chosen to work with the crystal
in the center of the simulation cell for convenience, but that the
results can be generalized to the case where the crystal straddles the
cell boundary \cite{zhang2016computing1,sayer2019finite}. The second
simulation geometry considered, the \emph{mirrored slab geometry}, is
shown in Fig.~\ref{fig:schematic}\,(c). In this case, one slab is
centered at $z_{\rm m}<(L-w)/2$, and its mirror image is centered at
$-z_{\rm m}$. The region between the slabs is occupied by an
electrolyte solution, while the regions exterior to the slabs (that
separate periodic images) are vacuum. In what follows we will
investigate the adsorption of ions from solution to polar surfaces in
these two common simulation geometries. We will demonstrate that the
mirrored slab geometry amounts to working with boundary conditions
that enforce $\sigma_\pm^{\rm (sim)}\approx 0$, a surely untenable
situation. In contrast, with an appropriate choice of boundary
conditions, we will show that the slab geometry yields
$\sigma_\pm^{\rm (sim)}\approx\sigma_\pm^{\rm (macro)}$, even with
relatively small simulation cells.

The remainder of the article is organized as follows. First, we will
present a brief overview of the \emph{finite field approach} and its
application to polar surfaces. This amounts to manipulating the
electrostatic boundary conditions. We will then present results for a
simple rocksalt system that demonstrates the severe implications that
the boundary conditions have on ion adsorption behavior. In
Sec.~\ref{sec:kao} we will show how this framework can be applied to
more complex systems, using kaolinite's basal surfaces as an
example. We will end with a summary and outlook for future directions.

\section{Controlling surface charge with finite fields}
\label{sec:ffm}

The difficulties in simulating systems like those shown in
Fig.~\ref{fig:schematic} originate from the long-ranged nature of
electrostatic interactions. In order to accurately compute Coulombic
forces, methods based on Ewald sums are typically used. Let
$\mathcal{H}_{\rm PBC}$ denote a Hamiltonian that includes
electrostatic interactions computed with an Ewald method (under the
so-called `tin foil boundary conditions'), along with any
non-electrostatic interactions. For the slab geometry, a natural
question arises: \emph{How quickly do simulations converge to the
  limit $L\to\infty$}?  With the implicit assumption that $w$ is held
fixed, it has long been known that the answer is ``not very''
\cite{spohr1997effect,YehBerkowitz1999sjc,yeh2011proper}. To avoid
large values of $L$ or computationally expensive two-dimensional
versions of Ewald sums, Yeh and Berkowitz (YB) devised a simple
correction scheme \cite{YehBerkowitz1999sjc} in which the system's
Hamiltonian is given by
\begin{equation}
  \mathcal{H}_{\rm YB} = \mathcal{H}_{\rm PBC} + 2\pi\Omega P^2,
\end{equation}
where $P$ is the $z$ component of the system's instantaneous
polarization, and $\Omega$ is the volume of the simulation cell. (In
our formulation, we use a unit system in which $4\pi\epsilon_0=1$,
where $\epsilon_0$ is the permittivity of free space.) The YB approach
has become one of the most widely used methods for simulating slab
systems. In the mirrored slab geometry, $\langle P\rangle \approx
0$. It could then be argued, at least on average, that the YB
correction term is redundant. Indeed, the mirrored slab geometry has
been proposed as a means to correct for unphysical long-ranged fields
arising from finite polar surfaces
\cite{croteau2009simulation,glatz2016surface,zielke2016simulations,glatz2017heterogeneous,ren2020effects,roudsari2019atomistic}. As
Fig.~\ref{fig:schematic} makes clear, however, a more pertinent
question for the current purpose is: \emph{How quickly do simulations
  converge to the limit where both $L$ and $w$ are macroscopic in
  extent}?

\begin{figure}[tb]
  \includegraphics[width=8cm]{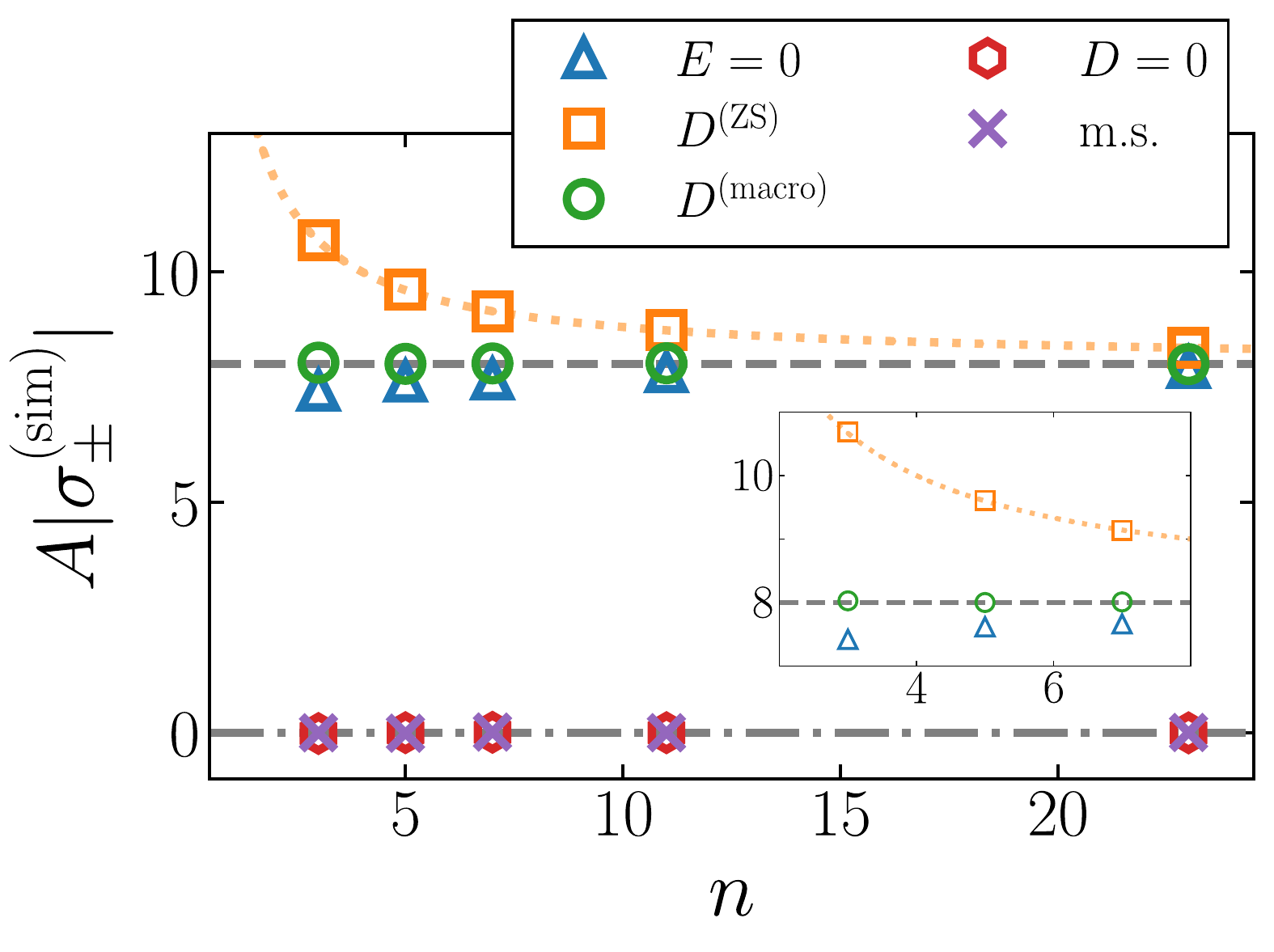}  
  \caption{Controlling $\sigma_\pm^{\rm (sim)}$ at rocksalt \hkl(111)
    through the electrostatic boundary conditions. The slab comprises
    $n+1$ layers with alternating charge density $\pm\sigma_0$ (see
    Fig.~\ref{fig:schematic}). The dashed line indicates
    $A|\sigma_\pm^{\rm (macro)}|=A\sigma_0/2$, where $A$ is the
    surface area. Using a standard Ewald approach ($E=0$) leads to
    slight underscreening for small $n$ (see inset). Using $D=0$ or
    the mirrored slab geometry (`m.s.') yields $|\sigma_\pm^{\rm
      (macro)}|\approx 0$ regardless of $n$. Using $D=D^{\rm (ZS)}$
    leads to significant overscreening at small $n$, but convergence
    to $|\sigma_\pm^{\rm (macro)}|$ is seen as $n$ increases. The
    dotted line indicates the theoretical prediction
    (Eq.~\ref{eqn:sigmaZS}). Using $D=D^{\rm (macro)}$ results in
    $|\sigma_\pm^{\rm (sim)}|\approx|\sigma_\pm^{\rm (macro)}|$ for
    all $n$.}
  \label{fig:sigmas}
\end{figure}

In Fig.~\ref{fig:sigmas} we show how $|\sigma_\pm^{\rm (sim)}|$ varies
for rocksalt \hkl(111), in contact with a concentrated aqueous NaCl
solution, as $n=w/R$ is varied. (The slab comprises $n+1$ layers
separated by a distance $R$.) Here we can see that when using
$\mathcal{H}_{\rm PBC}$, $|\sigma_\pm^{\rm (sim)}|$ approaches
$|\sigma_\pm^{\rm (macro)}|$ from below as $n$ increases. In other
words, for small $n$ the crystal's surface charge is
\emph{underscreened}. To fix this problem of underscreening, Zhang and
Sprik (ZS) proposed the application of an appropriate electric field
$E$ or electric displacement field $D$ across the simulation
cell. This is achieved with the finite field approach, as prescribed
by the Hamiltonian
\begin{equation}
  \mathcal{H}_E = \mathcal{H}_{\rm PBC} - \Omega EP\qquad\text{(constant $E$)}, \label{eqn:HE}
\end{equation}
or
\begin{equation}
  \mathcal{H}_D = \mathcal{H}_{\rm PBC} + \frac{\Omega}{8\pi}(D-4\pi P)^2\qquad\text{(constant $D$)}. \label{eqn:HD}
\end{equation}
In addition to charged interfacial systems
\cite{zhang2016finite,sayer2017charge,sayer2019finite,sayer2019stabilization,zhang2019coupling,zhang2018communication},
the finite field approach has been used to investigate the response of
both dielectrics
\cite{zhang2016computing1,zhang2016computing2,zhang2018note,zhang2020electromechanics,cox2020dielectric}
and ionic conductors \cite{pache2018molecular,cox2019finite}. For a
metal in contact with liquid electrolyte, it has also been shown to
give results indistinguishable from 2D Ewald
\cite{dufils2019simulating}. Moreover, although Eqs.~\ref{eqn:HE}
and~\ref{eqn:HD} were first derived on thermodynamic grounds
\cite{zhang2016computing1}, they are full microscopic Hamiltonians
that can also be derived from an extended Lagrangian based on
arguments of theoretical mechanics \cite{sprik2018finite}. See
Ref.~\onlinecite{zhang2020modelling} for a recent review. From
Eq.~\ref{eqn:HE}, it can be seen that using $\mathcal{H}_{\rm PBC}$ on
its own is equivalent to imposing $E=0$. ZS originally considered a
situation in which $n=1$ with increasing $R$, for which the rationale
of imposing an $E$ or $D$ field was simple; underscreening leads to an
erroneous electric field inside the slab, which is removed by imposing
an appropriate field. In this study, we will refer to these fields as
$E^{\rm (ZS)}$ and $D^{\rm (ZS)}$. While $E^{\rm (ZS)}$ is generally
found by trial-and-error, $D^{\rm (ZS)}$ can be obtained \emph{a
  priori} provided the structure of the crystal and $\sigma_0$ are
known. We will discuss this point in more detail below. Both $E^{\rm
  (ZS)}$ and $D^{\rm (ZS)}$ enforce the average field inside the slab
to vanish.

This approach was extended to polar crystal surfaces---specifically
rocksalt \hkl(111)---by Sayer \etal
\cite{sayer2017charge,sayer2019finite}, where $E^{\rm (ZS)}$ and
$D^{\rm (ZS)}$ were argued to decouple the two interfaces present in
the slab geometry [see Fig.~\ref{fig:schematic}\,(b)] such that the
double layer capacitance could be measured with small simulation
cells. A rather striking observation from
Refs.~\onlinecite{sayer2017charge,sayer2019finite}, however, is that
in the ZS approach the surface charge depends on $n$:
\begin{equation}
  \label{eqn:sigmaZS}
  \sigma_\pm^{\rm (sim,ZS)} = \frac{n+1}{n}\sigma_\pm^{\rm (macro)}.
\end{equation}
Thus, while $\lim_{n\to\infty}\sigma_\pm^{\rm
  (sim,ZS)}=\sigma_\pm^{\rm (macro)}$, significant deviations are
expected when $n$ is small. We mentioned above that $D^{\rm (ZS)}$ can
be established \emph{a priori} if certain properties of the system are
known. For rocksalt \hkl(111) \cite{sayer2017charge,sayer2019finite},
\begin{equation}
  \label{eqn:DZS}
  D^{\rm (ZS)} = 4\pi\frac{n+1}{n}\sigma_\pm^{\rm (macro)} \equiv
  4\pi\sigma_\pm^{\rm (sim,ZS)}.
\end{equation}
The choice of $\sigma_+^{\rm (macro)}$ or $\sigma_-^{\rm (macro)}$
depends upon the direction of the crystal's polarization $P_{\rm
  xtl}$, with $P_{\rm xtl}>0$ corresponding to $\sigma_{-}^{\rm
  (macro)}$ and \emph{vice versa}. The results for $|\sigma_\pm^{\rm
  (sim)}|$ obtained with $D=D^{\rm (ZS)}$ are shown in
Fig.~\ref{fig:sigmas}, along with the theoretical prediction given by
Eq.~\ref{eqn:sigmaZS}. As expected, $|\sigma_\pm^{\rm (macro)}|$ is
approached from above, and the results appear to be converging for
large $n$.

The ZS approach was designed as a means to compute the double layer
capacitance with relatively small simulation cells, and has enjoyed
success not only with classical force field models
\cite{zhang2016finite,sayer2017charge,sayer2019stabilization,zhang2018communication,dufils2019simulating},
but also with \emph{ab initio} approaches
\cite{sayer2019finite,zhang2019coupling}. Moreover, $n=1$ yields
$\sigma_{\pm}^{\rm (sim,ZS)} \approx \mp\sigma_0$, which is
appropriate if one is interested in modeling charged surfaces that
arise from e.g. protonation/deprotonation of surface groups. (See also
Ref.~\onlinecite{pan2019analytic} for an alternative approach for
tackling the $n=1$ system.) To obtain $\sigma_\pm^{\rm (sim)} \approx
\sigma_\pm^{\rm (macro)}$ for polar surfaces, however, it is clear
from Fig.~\ref{fig:sigmas} that a relatively large number of crystal
layers is required. This was the approach we adopted in
Ref.~\onlinecite{sayer2019stabilization} in our study of AgI in
contact with aqueous solution.\footnote{For the wurtzite crystal
  structure of AgI studied in
  Ref.~\onlinecite{sayer2019stabilization}, more rapid convergence
  with $n$ is seen than for the rocksalt structure.}  Should this be
necessary?  Eq.~\ref{eqn:DZS} suggests an inextricable link between
$D$ and $\sigma_\pm^{\rm (sim)}$:
\begin{center}
  \emph{The value of $D$ directly determines $\sigma_\pm^{\rm
      (sim)}$,\\ independent of $L$.}
\end{center}
This is the central message of this article. It is important to note
that implicit in this statement is that the electric field between
periodic replicas is assumed to vanish; this is ensured in our
simulations by the fact that the slab is surrounded by electrolyte.
While this relationship between $D$ and $\sigma_\pm^{\rm (sim)}$ can
be inferred from previous studies
\cite{zhang2016finite,sayer2019finite}---where derivations can also be
found---it has only been used to impose vanishing average electric
field inside the slab as a means to compute the double layer
capacitance. Here we provide empirical support showing this
relationship holds across a range of values for $D$, and demonstrate
its significance beyond calculating the double layer capacitance. From
this perspective, obtaining $\sigma_\pm^{\rm (sim)} \approx
\sigma_\pm^{\rm (macro)}$ is then a simple case of setting the
displacement field accordingly, i.e.,
\begin{equation}
  \label{eqn:Dmacro}
  D^{\rm (macro)} = 4\pi\sigma_\pm^{\rm (macro)}.
\end{equation}
Results from simulations with $D=D^{\rm (macro)}$ are shown in
Fig.~\ref{fig:sigmas}, where it is seen that $|\sigma_\pm^{\rm (sim)}|
\approx |\sigma_\pm^{\rm (macro)}|$ is an excellent approximation over
the range of $n$ investigated.

This relationship has a striking implication for the behavior of both
the YB approach and the mirrored slab geometry. It has been previously
noted that $\mathcal{H}_{\rm YB}$ and $\mathcal{H}_{D}$ with $D=0$ are
formally equivalent \cite{zhang2016finite}, and one might therefore
expect that $\sigma_\pm^{\rm (sim)}\approx 0$. Our results in
Fig.~\ref{fig:sigmas} confirm this notion and, along with the results
using $D=D^{(\rm ZS)}$ and $D=D^{\rm (macro)}$, provide convincing
evidence that the value of $D$ directly determines $\sigma_\pm^{\rm
  (sim)}$. For the mirrored slab geometry, $\mathcal{H}_{\rm PBC}$ on
its own is used. Recall that this corresponds to $E=0$ (see
Eq.~\ref{eqn:HE}). Thus, if $\langle P\rangle \approx 0$ then $\langle
D\rangle=E+4\pi \langle P\rangle \approx 0$, and the mirrored slab
geometry corresponds, on average, to $D=0$. We have previously shown
that for AgI \hkl(0001) in contact with pure water, the mirrored slab
geometry and regular slab geometry with $D=0$ give similar
electrostatic potential profiles, and orientation statistics for the
interfacial water molecules \cite{sayer2019stabilization}. For the
rocksalt \hkl(111) surface in contact with electrolyte considered
here, Fig.~\ref{fig:sigmas} shows that with the mirrored slab
geometry, $|\sigma_\pm^{\rm (sim)}|\approx 0$ over the range of $n$
considered. (The \ce{Cl-} planes are exposed to solution.) This
strongly suggests that both $\mathcal{H}_{\rm YB}$ and the mirrored
slab geometry are unsuitable for modeling systems like those depicted
in Fig.~\ref{fig:schematic}\,(a).

\section{Application to kaolinite's basal surfaces}
\label{sec:kao}

\begin{figure}[tb]
  \includegraphics[width=8cm]{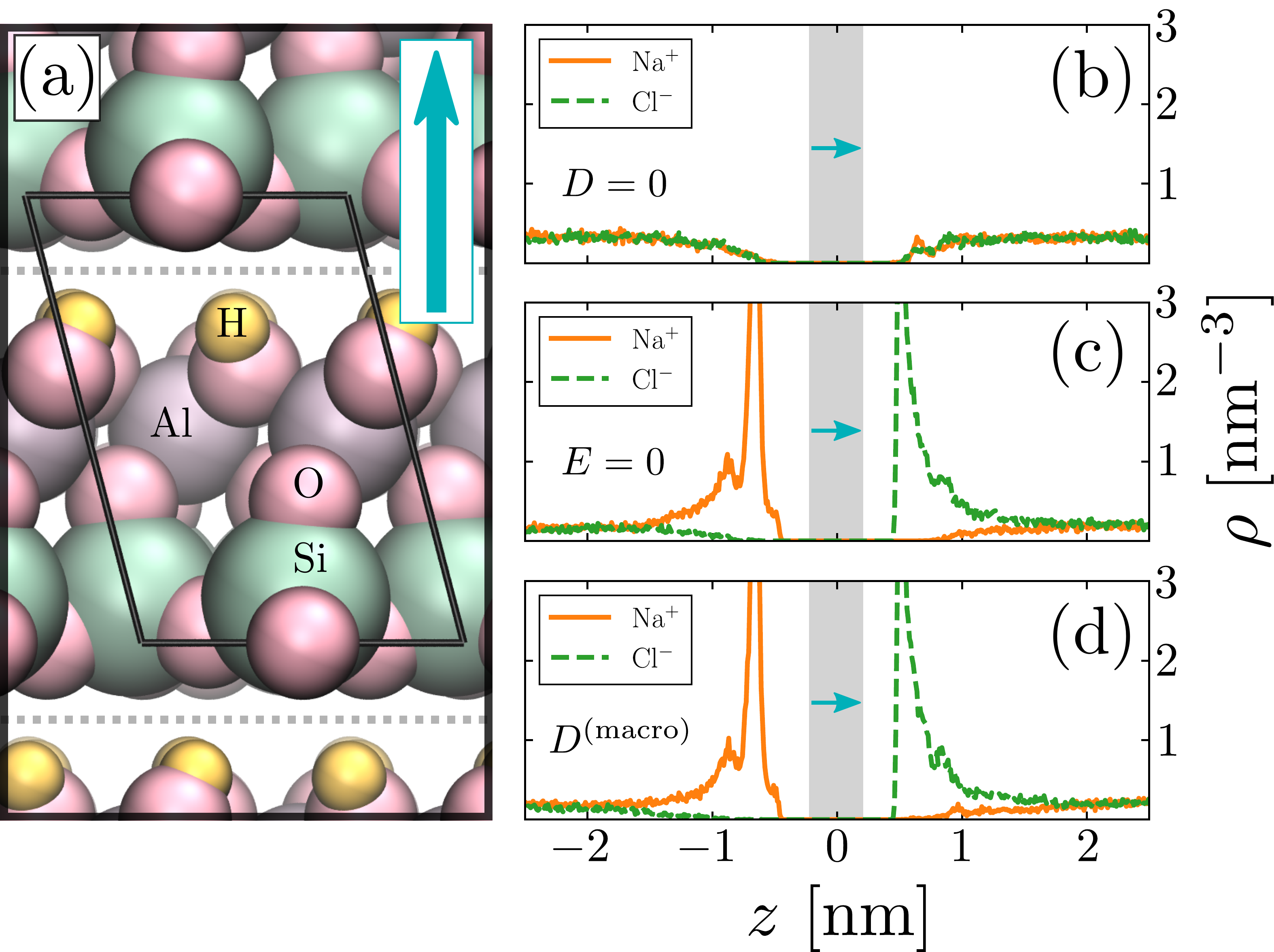}    
  \caption{Application to kaolinite's basal surfaces in contact with
    aqueous NaCl solution. (a) The bulk crystal structure of kaolinite
    comprises layers of \ce{Al2Si2O5(OH)4}. The basal surfaces are
    generated by cleaving relatively weak interlayer hydrogen bonds,
    as indicated by the gray dotted lines. The black lines delineate
    the bulk repeat unit. (b) Number density profiles $\rho$ of
    \ce{Na+} and \ce{Cl-} obtained at $D=0$. The gray shaded area
    approximately indicates the region occupied by kaolinite. The
    corresponding results obtained at $E=0$ and $D=D^{\rm (macro)}$
    are shown in (c) and (d), respectively. The blue arrows indicate
    the orientation of the crystal.}
  \label{fig:kao}
\end{figure}

The results presented so far demonstrate that $\sigma_\pm^{\rm
  (macro)}$ can be obtained for any value of $n$ provided one uses the
appropriate electrostatic boundary conditions. This is achieved most
straightforwardly by setting $D=4\pi\sigma_\pm^{\rm (macro)}$ in
$\mathcal{H}_D$ (Eq.~\ref{eqn:HD}). So far, we have only tackled the
relatively simple rocksalt \hkl(111) surface. Here we demonstrate the
relevance of the principles established in Sec.~\ref{sec:ffm} to a
more complex system, namely the basal surfaces of kaolinite, an
aluminosilicate clay mineral. These surfaces are widely studied with
molecular simulation owing to their importance in ice nucleation and
geochemistry
\cite{cox2013microscopic,sosso2016ice,zielke2015simulations,glatz2017heterogeneous,cox2018formation,vasconcelos2007molecular,tenney2014molecular}. To
proceed, we need to establish an estimate for $\sigma_\pm^{\rm
  (macro)}$ for the crystal structure shown in
Fig~\ref{fig:kao}\,(a). To this end, we will simply use established
results from the solid state community. For a detailed discussion of
the underlying theory, we refer the reader to the review by
Goniakowski \etal \cite{goniakowski2008polarity}

While kaolinite presents a complex crystal structure, an estimate for
$\sigma_\pm^{\rm (macro)}$ can in fact be determined in a rather
simple fashion, and furthermore highlights an essential aspect of the
theory of polar surfaces: \emph{The dipole moment $\mu_{\rm B}$ of the
  bulk repeat unit determines $\sigma_\pm^{\rm (macro)}$}. This means
that $\sigma_\pm^{\rm (macro)}$ may depend upon how the bulk crystal
structure is terminated, which can be important for materials such as
\ce{TiO2}, \ce{Al2O3} and \ce{SrTiO3}. For kaolinite, however, it is
natural to cleave its basal surfaces such that only relatively weak
hydrogen bonds are broken, as indicated by the gray dotted line in
Fig.~\ref{fig:kao}\,(a). With the \textsc{CLAYFF} force field
\cite{cygan2004molecular} used in this study, we find $|\mu_{\rm B}|
\approx 14.1$\,D. For comparison, the rocksalt \hkl(111) surface with
the Joung-Cheatham force field \cite{joung2008determination} gives
$|\mu_{\rm B}| \approx 7.8$\,D. Denoting the volume of the repeat unit
as $\Omega_0$, our estimate for $\sigma_\pm^{\rm (macro)}$ is then
simply given by
\begin{equation}
  \label{eqn:muBSigma}
  \sigma_\pm^{\rm (macro)} = \frac{\mp|\mu_{\rm B}|}{\Omega_0}.
\end{equation}
It is straightforward to verify that for rocksalt \hkl(111),
Eq.~\ref{eqn:muBSigma} recovers $\sigma_\pm^{\rm
  (macro)}=\mp\sigma_0/2 \approx\mp 3.6$\,$e$/nm$^2$. For kaolinite we
find $\sigma_\pm^{\rm (macro)}\approx\mp 0.89$\,$e$/nm$^2$. While we
therefore expect quantitative differences between rocksalt \hkl(111)
and kaolinite's basal surfaces, we nonetheless expect a comparable
(i.e., same order of magnitude) coverage of adsorbed counterions at
the two surfaces. We note in passing that Eq.~\ref{eqn:muBSigma}
states that $\sigma_\pm^{\rm (macro)}$ is determined by properties of
the bulk repeat unit, and is not related to any surface dipole that
may exist.

Performing simulations for a single sheet of kaolinite with its basal
surfaces in contact with aqueous solution, and its atoms fixed in
their bulk crystal lattice positions, corroborates the findings
presented in Sec.~\ref{sec:ffm}: \tcr{$|\sigma_\pm^{\rm (sim)}|
  \approx 0.89$\,$e$/nm$^2$} using $\mathcal{H}_E$ and $E=0$;
\tcr{$|\sigma_\pm^{\rm (sim)}| \approx 0.00$\,$e$/nm$^2$} using
$\mathcal{H}_D$ and $D=0$; and \tcr{$|\sigma_\pm^{\rm (sim)}| \approx
  |\sigma_\pm^{\rm (macro)}| \approx 0.89$\,$e$/nm$^2$} using
$\mathcal{H}_D$ and $D=D^{\rm (macro)}$ as given by
Eqs.~\ref{eqn:Dmacro} and~\ref{eqn:muBSigma}. Perhaps more striking,
however, are the ion density profiles, as shown in
Figs.~\ref{fig:kao}\,(b)-(d). Here we see that simulations using $D=0$
give \emph{qualitatively incorrect} results, with essentially no ion
adsorption observed. This result is broadly in line with Ren \etal,
who used the mirrored slab geometry, and even reported slightly more
favorable adsorption of cations vs. anions at kaolinite's positive
\hkl(0001) surface \cite{ren2020effects}. (Results from the mirrored
slab geometry are shown in \tcr{Fig.~\ref{fig:kao-mirror}}, and are in
good agreement with those from simulations using $D=0$.) In contrast,
with both $E=0$ and $D=D^{\rm (macro)}$ we see behavior in line with
physical intuition, with \ce{Na+} and \ce{Cl-} ions adsorbed to the
negative \hkl(000-1) and positive \hkl(0001) surfaces,
respectively. These results are also consistent with Vasconcelos \etal
\cite{vasconcelos2007molecular}, who investigated ion adsorption at
kaolinite's basal faces with a standard Ewald method. Moreover,
performing simulations with $E=0$ for two and three sheets of
kaolinite yields \tcr{$|\sigma_\pm^{\rm (sim)}| \approx
  0.88$\,$e$/nm$^2$} in both cases suggesting that, even with a single
sheet of kaolinite, results are sufficiently converged. On the whole,
for both the rocksalt \hkl(111) and kaolinite systems, we find $E=0$
simulations yield largely satisfactory results. How quickly results
converge as $n$ increases, however, appears to be system-dependent.

Throughout this article, we have deliberately avoided detailed
theoretical discussions, instead choosing to focus on empirically
demonstrating how different electrostatic boundary conditions affect
$\sigma_\pm^{\rm (macro)}$. Nonetheless, we end this section with a
couple of comments concerning the underlying theory. First, the
relation $D=4\pi\sigma_\pm^{\rm (sim)}$ can be interpreted as a
statement that the `virtual electrodes' directly influence the
behavior of the system at the cell boundaries (see
Refs.~\onlinecite{zhang2016computing1,zhang2016finite}). In the case
that an electrolyte---which has unit polarizability---straddles the
cell boundary, its polarization is then immediately determined: $4\pi
P = D$. The surface charge density in the double layer then follows
from basic electrostatic arguments i.e., $4\pi\sigma_\pm^{\rm
  (sim)}=4\pi P = D$. Crucial to this argument is that the ions are
included in the polarization. Second, if $D=D^{\rm (ZS)}$ enforces a
vanishing average electric field inside the crystal, what is the
effect of $D=D^{\rm (macro)}$? Enforcing $\sigma_\pm^{\rm (sim)} =
\sigma_\pm^{\rm (macro)}$ removes the linear component of the
electrostatic potential $\phi(z)$ in the crystal's interior e.g. in
the case of rocksalt \hkl(111), $\phi(z) = \phi(z+2R)$, whereas
$D=D^{\rm (ZS)}$ imposes $\phi(-w/2) = \phi(w/2)$. Thus while with
$D=D^{\rm (macro)}$ an electrostatic potential difference across the
crystal remains, it does not grow with $w$, and avoids the so-called
`polar-catastrophe'
\cite{tasker1979stability,nosker1970polar,goniakowski2008polarity,noguera2000polar}.

\begin{figure}[b]
  \includegraphics[width=8cm]{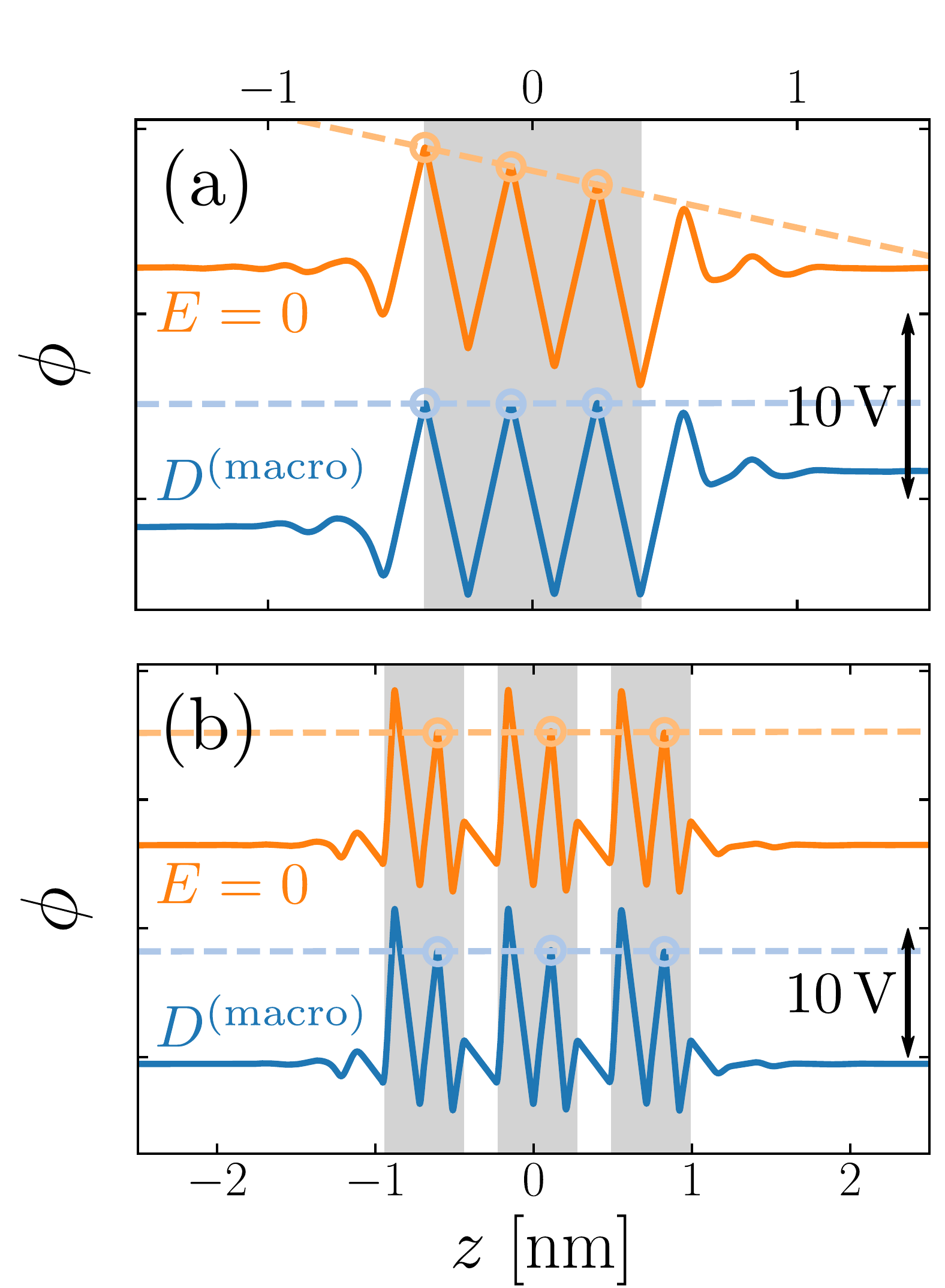}    
  \caption{Electrostatic potential profiles $\phi$ for (a) rocksalt
    \hkl(111) with $n=5$ and (b) three sheets of kaolinite. In (a),
    circles indicate $\phi$ evaluated at planes of \ce{Na+} ions,
    which for $E=0$ decreases linearly across the crystal. The dashed
    line indicates a linear fit. Imposing $D=D^{\rm (macro)}$ removes
    this linear contribution. The gray shaded area indicates the
    region occupied by the crystal. In (b), circles indicate $\phi$
    evaluated at planes of octahedral aluminum sites. For both $E=0$
    and $D=D^{\rm (macro)}$, negligible linear component in $\phi$ is
    seen in the bulk of the crystal. The gray shaded areas indicate
    the regions occupied by the kaolinite sheets, and their
    orientation is the same as in Fig.\,\ref{fig:kao}.}
  \label{fig:phis}
\end{figure}

In Fig.~\ref{fig:phis} we present $\phi(z)$ \cite{wirnsberger2016non}
for both the rocksalt \hkl(111) system with $n=5$, and the kaolinite
system with three sheets of crystal. In the case of the former
[Fig.~\ref{fig:phis}\,(a)], $\phi$ exhibits a significant linear
component within the crystal's interior, which is indeed removed by
imposing $D=D^{\rm (macro)}$. In contrast, for kaolinite
[Fig.~\ref{fig:phis}\,(b)] we see that $\phi$ is broadly similar
between $E=0$ and $D=D^{\rm (macro)}$, which is reflected in the
similar values for $\sigma_\pm^{\rm (sim)}$ reported
above. Importantly, negligible linear component in $\phi$ is observed,
giving us confidence that Eq.~\ref{eqn:muBSigma} provides a good
estimate for $\sigma_\pm^{\rm (macro)}$, even for complex systems like
kaolinite.

\section{Summary and Outlook}

In this article, we have investigated the effect of different
electrostatic boundary conditions on simulated observables such as ion
distributions and integrated surface charge densities for polar
crystal surfaces in contact with aqueous solution. We have shown that
on average, the mirrored slab geometry with $E=0$ and slab geometry
with $D=0$ give similar, but intuitively incorrect,
results. Specifically, such simulation conditions impose a vanishing
integrated surface charge density. Using results from studies on polar
surfaces by the solid state community \cite{goniakowski2008polarity},
combined with recent developments in performing molecular dynamics
simulations at constant $E$ and $D$
\cite{zhang2016computing1,zhang2016finite}, we have shown that one can
obtain sensible surface charge densities with relatively small
simulation cells. We also showed how this approach can be applied to
complex systems such as clay minerals. Although we have previously
demonstrated the use of the finite field approach for ice formation at
AgI's polar surfaces \cite{sayer2019stabilization}, for liquid/solid
interfaces they have primarily been used as a tool to compute the
double layer capacitance. While undoubtedly an important property,
what this work makes clear is that this framework also provides a
means to understand the effects of electrostatic boundary conditions
on simulated observables of general importance, such as average
structural properties.

We are of course ultimately interested in `correct' rather than
`sensible' results. Neglecting issues concerning the underlying simple
point charge force fields (including their appropriateness for
calculating the bulk polarization in Eq.~\ref{eqn:muBSigma}, see
e.g. Ref.~\onlinecite{jiang2012rigorous}), those presented here should
be a good approximation for polar crystal surfaces with a
bulk-terminated crystal structure, and where all polarity compensation
arises by adsorption of ions from solution. Allowing for surface
relaxation will likely manifest itself as a relatively small
perturbation \cite{sayer2019stabilization}. In contrast, ascertaining
the relative importance of different polarity compensation
mechanisms---such as non-stoichiometric or electronic
reconstruction---remains an open and challenging question. Addressing
this issue will be a key step in establishing what the stable
structures of polar crystal surfaces actually are in a solution
environment. This will likely be important in the future development
of crystal structure prediction approaches as they try to incorporate
more information regarding the influence of the solution environment
\cite{price2018zeroth}.

\section{Methods}
\label{sec:methods}

All simulations used the SPC/E water model
\cite{BerendsenStraatsma1987sjc}, whose geometry was constrained using
the RATTLE algorithm \cite{andersen1983rattle} and the Joung-Cheatham
NaCl force field \cite{joung2008determination}. For simulations
involving kaolinite, the CLAYFF force field was used
\cite{cygan2004molecular}. Lorentz-Berthelot mixing rules were used to
compute Lennard-Jones interactions between different species. Dynamics
were propagated using the velocity Verlet algorithm with a time step
of 2\,fs. The temperature was maintained at 298\,K with a
Nos\`{e}-Hoover chain \cite{shinoda2004rapid,tuckerman2006liouville},
with a damping constant 0.2\,ps. The particle-particle particle-mesh
Ewald method was used to account for long-ranged interactions
\cite{HockneyEastwood1988sjc}, with parameters chosen such that the
root mean square error in the forces were a factor $10^{5}$ smaller
than the force between two unit charges separated by a distance of
0.1\,nm \cite{kolafa1992cutoff}. A cutoff of 1\,nm was used for
non-electrostatic interactions. For results in the main article, the
LAMMPS simulation package was used throughout
\cite{plimpton1995sjc}. For simulations with a $D$ field, the
implementation given in Ref.~\citenum{cox2019finite} was used. Results
from simulations using the GROMACS 4 simulation package
\cite{hess2008gromacs} are presented in
\tcr{Fig.~\ref{fig:sigmas_gromacs}}.

For the results presented in Fig.~\ref{fig:sigmas}, the electrolyte
comprised 600 water molecules and 20 NaCl ion pairs. The crystal
consisted of alternating layers of \ce{Na+} and \ce{Cl-} ions,
separated by $R=0.1628$\,nm, and each layer comprised 16 ions. The
lateral dimensions of the simulation cell were $L_x=1.5952$\,nm and
$L_y=1.3815$\,nm along $x$ and $y$, respectively. In the slab geometry
with $n=3$, the length of the simulation cell along $z$ was
$L=9.4841$\,nm, and $L$ was increased with $n$ accordingly e.g. for
$n=5$, $L$ was increased by $2R$. For the mirrored slab geometry, $L$
was double that of the corresponding simulation in the slab
geometry. Each simulation was 10\,ns long post equilibration.

For results presented in Fig.~\ref{fig:kao}, the electrolyte comprised
605 water molecules and 5 NaCl ion pairs. The bulk kaolinite structure
was taken from Ref.~\onlinecite{tenney2014molecular}. An orthorhombic
simulation cell was used with $L_x=1.5462$\,nm and
$L_y=1.7884$\,nm. For simulations with a single sheet $L=7.5$\,nm,
while for simulations with two and three sheets, $L=8.2162$\,nm and
$8.9323$\,nm, respectively. Simulations were 100\,ns long post
equilibration. Results presented in Fig.~\ref{fig:phis}\,(b) used the
same settings, except simulations were 20\,ns long post equilibration.

\section*{Supplementary Material}
Supplementary Material includes results from simulations for kaolinite
in a mirrored slab geometry, along with results for rocksalt \hkl(111)
using the GROMACS simulation package.

\begin{acknowledgments}
  Michiel Sprik and Chao Zhang are thanked for their many insights on
  this topic. We are grateful for computational support from the UK
  Materials and Molecular Modelling Hub, which is partially funded by
  EPSRC (EP/P020194), for which access was obtained via the UKCP
  consortium and funded by EPSRC grant ref EP/P022561/1. T.S. is
  supported by a departmental studentship (No. RG84040) sponsored by
  EPSRC. S.J.C. is supported by a Royal Commission for the Exhibition
  of 1851 Research Fellowship.
\end{acknowledgments}

\section*{Data Availability Statement}

The data that supports the findings of this study are available within
the article and its supplementary material. Input files for the
simulations are openly available at the University of Cambridge Data
Repository, \url{https://doi.org/10.17863/CAM.56629}.

\bibliography{./cox}

\clearpage
\onecolumngrid
\renewcommand\thefigure{S\arabic{figure}}
\renewcommand\theequation{S\arabic{equation}}
\setcounter{figure}{0}
\setcounter{equation}{0}

\noindent {\Large \tbf{Supplementary Material}}

\section*{Kaolinite in a mirrored slab geometry}

\begin{figure}[htbp]
  \includegraphics[width=12cm]{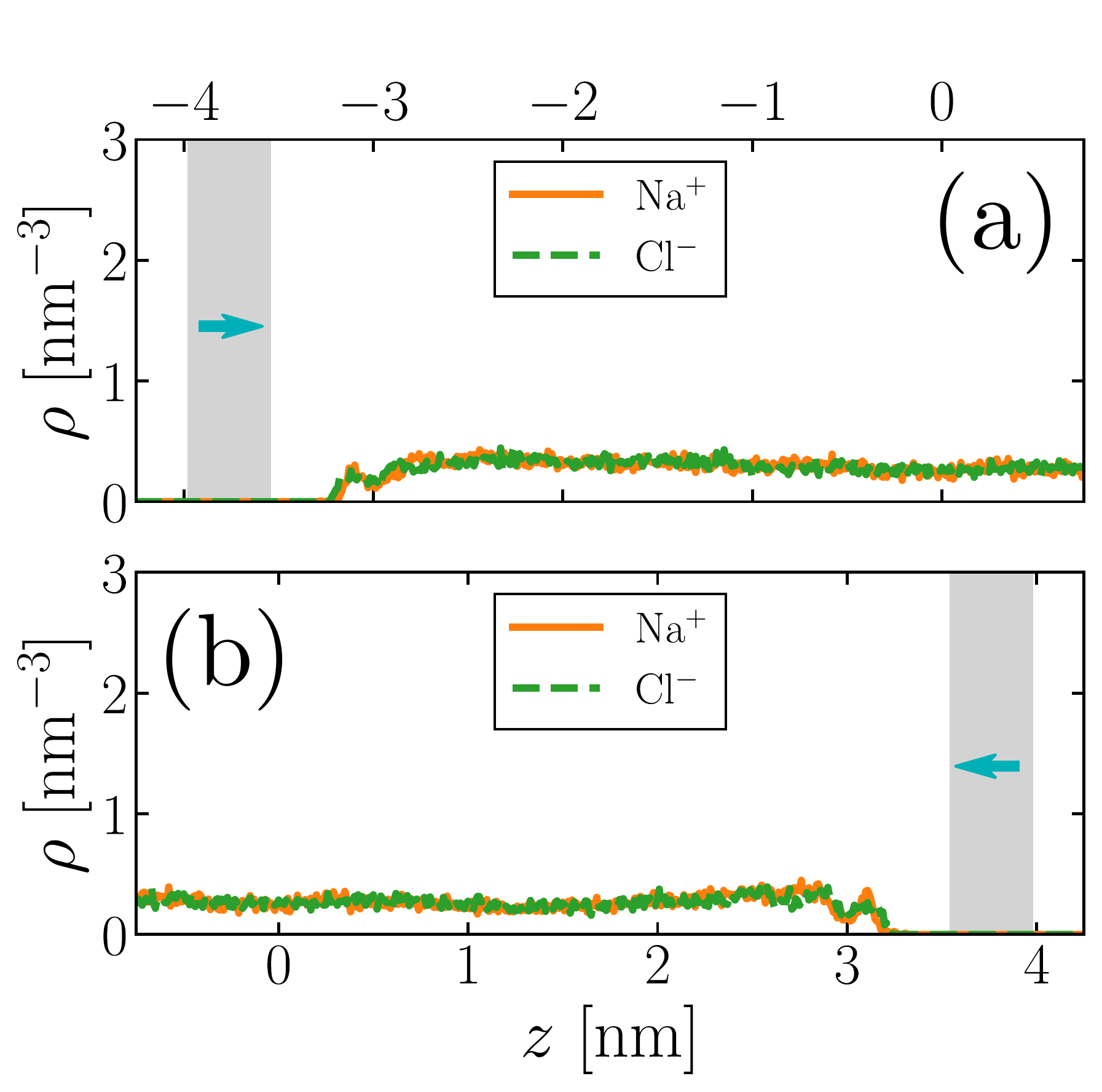}
  \caption{Number density profiles $\rho$ of \ce{Na+} and \ce{Cl-} for
    kaolinite \hkl(0001) in contact with aqueous electrolyte solution,
    in a mirrored slab geometry. (a) and (b) focus on the slabs at
    $z=-z_{\rm m}$ and $z=+z_{\rm m}$, respectively. The blue arrows
    indicates the orientation of the crystal [see
      Fig.~\ref{fig:kao}\,(a)]. These results agree well with those at
    $D=0$ [see Fig.~\ref{fig:kao}\,(b)]. Simulation settings were the
    same as described in the main text, with $L=15.0$\,nm. We find
    $\sigma_+^{\rm (sim)}\approx 0.00$\,$e$/nm$^2$.}
  \label{fig:kao-mirror}
\end{figure}

\clearpage

\section*{Results from GROMACS}

\begin{figure}[htbp]
  \includegraphics[width=12cm]{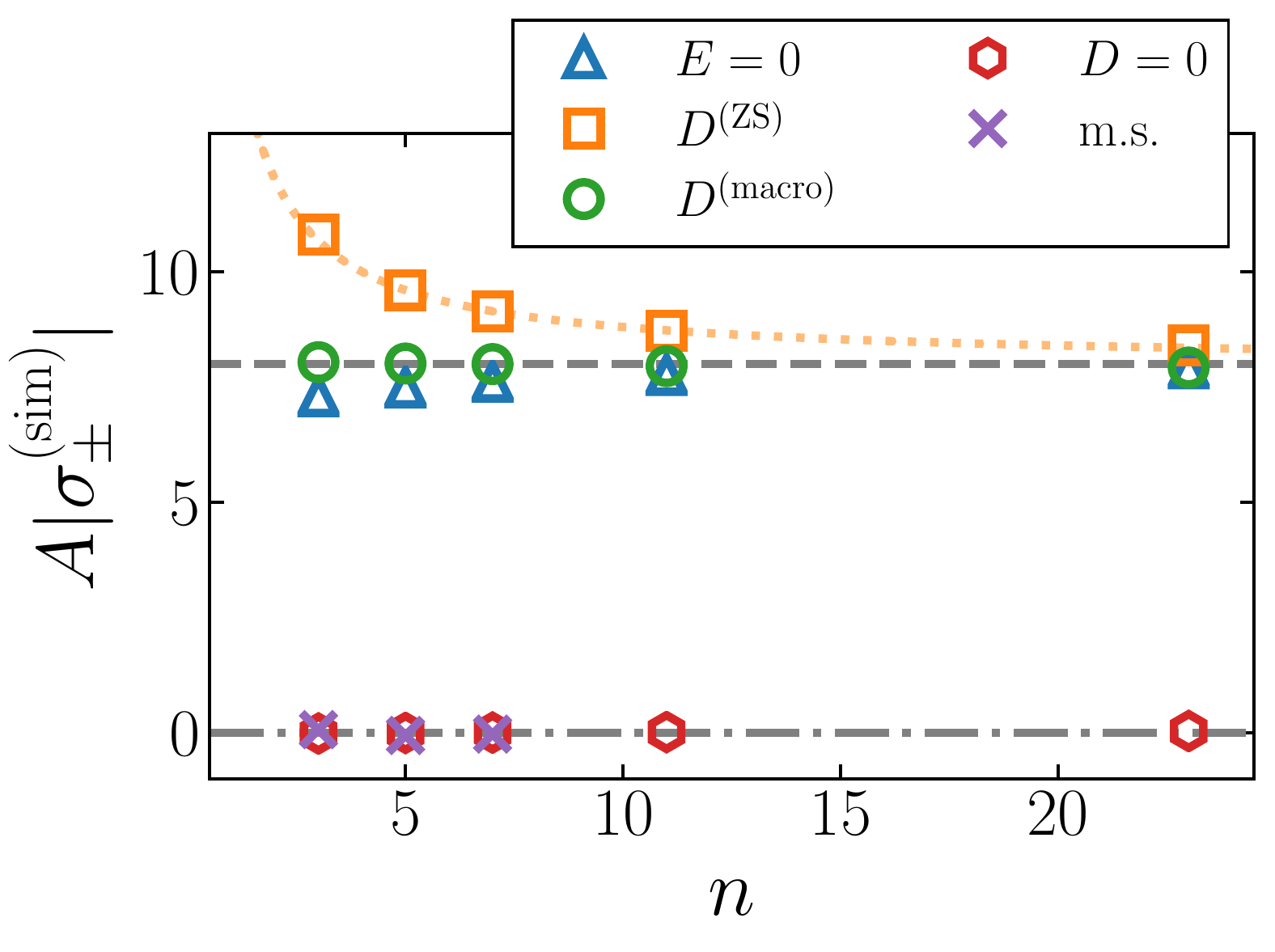}
  \caption{Results for the rocksalt \hkl(111) system obtained with
    GROMACS. These are in excellent agreement with results obtained
    with LAMMPS (see Fig.~\ref{fig:sigmas}).}
  \label{fig:sigmas_gromacs}
\end{figure}

We have also performed simulations of the rocksalt \hkl(111) system
using the GROMACS 4 simulation package
\cite{hess2008gromacs}. Simulation settings are broadly similar to
those described in the main text, and specific details can be found in
Refs.~\onlinecite{sayer2017charge}
and~\onlinecite{sayer2019finite}. For the mirrored slab geometry
(\ce{Na+} exposed) a 3\,nm vacuum gap was employed.  All simulations
were at least 2\,ns long. Results are presented in
Fig.~\ref{fig:sigmas_gromacs}, and are in excellent agreement with
those obtained with LAMMPS (see Fig.~\ref{fig:sigmas}). It is worth
noting that the implementation of $\mathcal{H}_D$ in GROMACS was
performed independently (see Ref.~\onlinecite{zhang2016computing1})
from the implementation in LAMMPS (see
Refs.~\onlinecite{cox2019finite}
and~\onlinecite{sayer2019stabilization}).

\end{document}